\documentclass[structabstract]{aa}   
\usepackage{graphicx} 
\usepackage{txfonts} 
\usepackage{natbib} 
\bibpunct{(}{)}{;}{a}{}{,} % to follow the A&A style 
\newcommand{\be}{\begin{displaymath}} 
\newcommand{\ee}{\end{displaymath}} 
\newcommand{\beq}{\begin{equation}} 
\newcommand{\eeq}{\end{equation}} 
\begin{document} 
  	\title{Phase resolved spectroscopic study of the isolated neutron star
          RBS 1223 (1RXS\,J130848.6+212708) \thanks{Based on observations obtained with 
 {\em XMM-Newton}, an ESA science mission with instruments and contributions directly funded by ESA Member States and NASA} }

   \author{V. Hambaryan 
          \inst{1} 
%          \inst{1}\fnmsep\thanks{Valeri Hambaryan, e-mail:~vvh@astro.uni-jena.de} 
\and
	  V. Suleimanov 
          \inst{2,3} 
	  \and 
	  A.D. Schwope 
          \inst{4} 
          \and 
	  R. Neuh\"auser 
          \inst{1} 
          \and 
	  K. Werner 
          \inst{2} 
	  \and 
	  A.Y. Potekhin 
	  \inst{5,6,7} 
          } 
 
   \institute{Astrophysikalisches Institut und Universit\"ats-Sternwarte,  
           Universit\"at Jena, Schillerg\"a\ss chen 2-3,  
           07745 Jena, Germany\\ 
           \email{vvh@astro.uni-jena.de} 
         \and 
             Institute for Astronomy and Astrophysics, 
           Kepler Center for Astro and Particle Physics, 
           Eberhard Karls University, 
           Sand 1, 
           72076 T\"ubingen, 
           Germany, 
      \and 
           Kazan  Federal University, 
           Kremlevskaja Str., 18, 
           Kazan 420008, 
           Russia 
	\and 
	   Leibniz-Institut f\"ur Astrophysik Potsdam,  
           An der Sternwarte 16,  
           14482 Potsdam, Germany 
	\and 
	 Ioffe Physical-Technical Institute,   
         Politekhnicheskaya Str., 26,  
         St.Petersburg 194021, Russia   
	\and 
	CRAL, 
	Ecole Normale Sup\'{e}rieure de Lyon,
	UMR CNRS No. 5574,
	Universit\'e de Lyon,
	69364 Lyon Cedex 07, France
	\and 
	Isaac Newton Institute of Chile, St.~Petersburg Branch, Russia 
             } 
 
   \date{Received ... / Accepted ...} 
 
% \abstract{}{}{}{}{}  
% 5 {} token are mandatory 
  
  \abstract 
	{} 
	{To constrain the mass-to-radius ratio of isolated neutron stars,  
         spin-phase resolved X-ray spectroscopic analysis is performed.} 
{The data from all observations of RBS 1223 (1RXS\,J130848.6+212708) conducted by 
\emph{XMM-Newton} EPIC pn  with the same instrumental setup in 2003-2007 were combined 
to form  spin-phase resolved spectra. A number of complex models of neutron stars 
with strongly magnetized  
($B_\mathrm{pole} \sim 10^{13}-10^{14}$~G) surface,  
    with temperature and magnetic field distributions around magnetic poles,  
    and partially ionized hydrogen thin atmosphere above it  
    have been implemented into the X-ray spectral fitting package \emph{XSPEC} for 
    simultaneous fitting of phase-resolved spectra.  
    A Markov-Chain-Monte-Carlo (MCMC) approach is also applied to verify 
    results of fitting and estimating in multi parameter models.} 
    {The spectra in different rotational phase intervals and light curves 
    in different energy bands with high S/N ratio show a high complexity. 
    The spectra can be parameterized with a Gaussian absorption line superimposed on a 
    blackbody spectrum, while the light curves with double-humped shape  
    show strong dependence of pulsed fraction upon the energy band (13\%\,--\,42\%),  
    which indicates that radiation emerges from at least two emitting areas.} 
    {A model with condensed iron surface and  partially ionized hydrogen 
     thin atmosphere above it allows us to fit simultaneously the observed 
     general spectral shape and the broad absorption feature observed at 0.3 keV 
     in different spin phases of RBS 1223. 
    It allowed to constrain some physical properties of X-ray emitting areas, i.e. 
    the temperatures  
     ($T_\mathrm{p1}\sim 105$~eV, $T_\mathrm{p2}\sim 99$~eV), magnetic field strengths  
    ($B_\mathrm{p1}\approx B_\mathrm{p2} \sim 8.6\times10^{13}$~G) at the poles, and  
    their distributions parameters ($a_\mathrm{1}\sim 0.61, a_\mathrm{2}\sim 0.29$, indicating  
    an absence of strong toroidal magnetic field component).  
    In addition, it puts some constraints on the geometry of the emerging X-ray emission  
    and gravitational redshift ($z=0.16^{+0.03}_{-0.01}$) of RBS 1223.}
 
   \keywords{stars: individual: RBS 1223 (1RXS\,J130848.6+212708) --
   stars: neutron -- stars: atmospheres -- X-rays: stars} 
 
   \titlerunning{Phase resolved spectroscopy of RBS 1223}

   \maketitle 
% 
%________________________________________________________________ 
 
\section{Introduction} 
 
   The study of thermally emitting, radio-quiet, nearby, isolated neutron stars  
(INSs) may have an important impact on our understanding of the physics of neutron stars.  
Observations and modeling of thermal emission from INSs can provide not only 
information on the physical properties such as the magnetic field, temperature, 
and chemical composition of the regions where this radiation is produced, but also 
information on the properties of matter at higher densities deeper inside the star. 
 
In particular, measuring the gravitational redshift of an identified spectral feature   
in the spectrum of thermal radiation emitted from the INS surface/atmosphere  
may provide a useful constraint on theoretical models of equations of state  
for  superdense matter. Independent of the estimate of the INS  
radius \citep[e.g.][]{2004NuPhS.132..560T}  
from the thermal spectrum of an INS, it may allow us to 
directly estimate the mass-to-radius ratio.   
 
Much more information may be extracted by studying spin-phase resolved  
spectra with high signal to noise ratio and  
fitting them with model spectra of radiation
emitted from highly magnetized INS surface layers.  
 
\object{RBS 1223} (1RXS\,J130848.6+212708) was originally discovered as a soft X-ray source during the 
\emph{ROSAT} All-Sky Survey by \cite{1999A&A...341L..51S}.  
It shared common characteristics with
other members of the small group (so far 7 discovered by \emph{ROSAT})
of thermally emitting and radio-quiet, nearby INSs, traditionally dubbed 
XDINS (from ``X-ray dim INS'') or Magnificent Seven
(see, e.g., reviews by
\citealp{2007Ap&SS.308..181H,2008A&ARv..15..225M,2009ASSL..357..141T},
and references therein): 
soft spectra, well described by blackbody radiation  
with temperatures  60\,--\,120~eV, no other spectral features, no association with SNR,  
no radio emission, no X-ray pulsations, very high X-ray to optical flux 
ratio. A nature of these sources as old INS reheated by accretion from the 
interstellar medium or young cooling stars seemed possible.  
Meanwhile, intensive X-ray observations with mainly  \emph{XMM-Newton} and also with \emph{Chandra},  
as well as optical/UV observations of most probable counterparts, have revised this 
picture in parts and has provided intriguing physical insight. 
 
X-ray pulsations have been found in six of these objects,
with periods clustered at 3\,--\,11~sec, 
and period derivatives have been measured for five of them. In a classic $P-\dot{P}$ 
diagram the XDINSs are found intermediate between radio pulsars 
and magnetars \citep[e.g.,][]{2008A&ARv..15..225M}.  
The inferred magnetic field strengths are above $10^{13}$ G. 
 
Imaging CCD-spectroscopy with \emph{XMM-Newton}
has uncovered absorption features in 
at least three, likely in six stars. At current energy resolution 
($\textrm{FWHM} \sim 50$\,--\,150 eV 
between $0.2$\,--\,2.0 keV) they can be, formally, well fitted as a Gaussian absorption lines,
which are usually connected with ion cyclotron lines.
Their interpretation is not unique, magnetically shifted atomic transitions or 
electron cyclotron resonances are debated.  
For some of the models, 
the inferred magnetic field strengths are again 
above $10^{13}$ G \citep[][]{2007Ap&SS.308..181H}.  
 
Among them, RBS 1223 is a special case, some sort of outlier,  
qualitatively different from the typical pattern observed in the sample of XDINSs. 
Namely, the  rotational phase folded light curve  has a double-humped shape,  
with largest pulsed fraction\footnote{The pulsed fraction, used in this paper, is defined as:  
\be 
\mathrm{PF} \equiv \frac{\mathrm{CR}_\mathrm{max}-\mathrm{CR}_\mathrm{min}
     }{
     \mathrm{CR}_\mathrm{max}+\mathrm{CR}_\mathrm{min}}, 
\ee 
where CR is the count rate. Note that this definition of a pulsed fraction may be misleading 
in the case of complex shaped light curves,  
a peak/minimum flux may appear at different phases in different energy ranges.  
Instead, the following quantity, i.e semi-amplitude of modulation, might be an  
appropriate descriptor of the pulsed emission: 
\be 
A \equiv \frac{\sum_i{|\mathrm{CR}_{i}-\langle \mathrm{CR}_{i}\rangle|}}{\sum_i{\mathrm{CR}_{i}}}, 
\ee 
where $\mathrm{CR}_{i}$ is a count rate per phase bin.} 
($\sim$~19\% in 0.2-1.2 keV energy range, see further).  
Moreover, the separation of maxima by less than 180 degrees, significantly  
different count-rates of minima, and the variation of the blackbody apparent 
temperatures between 80 and 90 eV over the spin cycle are evidence 
for, at least, two emitting areas with some temperature distribution over neutron 
star surface \citep{2005A&A...441..597S,2007Ap&SS.308..619S}. 
 
It is worth to note that the abovementioned observed absorption feature at $\sim$~0.3 keV in  
the spectrum of RBS 1223 has the highest equivalent width 
\citep[$\sim$~0.2keV,][]{2007Ap&SS.308..619S} among all XDINSs. Moreover, spectral analysis of the average  
XMM-Newton spectrum of RBS 1223 \citep[][]{2007Ap&SS.308..619S}  
based on the two first observations (see Table~\ref{obslog}) showed the possible  
presence of a second feature in the X-ray spectrum.  
Its existence, however, is not uniquely proven because of some remaining 
calibration uncertainties and the not well-defined continuum at high energies 
because of the lack of photons (lower S/N ratio).  
 
Meanwhile, new observational sets (see Table~\ref{obslog}) are publicly available  
and a number of new detailed models of emergent spectra of highly magnetized INSs  
are developed and available \citep{2009AAS...21343609H,2010A&A...522A.111S}. 
 
Preliminary analysis of the phase averaged spectrum of RBS 1223 
was performed by \cite{2006A&A...459..175P}. They mentioned  that there 
is a possibility of good fits with quadrupolar magnetic fields, although not 
excluding a condensed surface model with hydrogen atmosphere,  
including vacuum polarization effects \citep{2006MNRAS.373.1495V}. 
 
To explain unusual observed properties of RBS 1223 
Suleimanov et al. (2010a, hereafter Paper I) 
studied various local models of the
emitting surface of this INS. They considered three types of models:  
naked condensed surfaces,
semi-infinite partially ionized hydrogen atmospheres with vacuum
polarization and partial mode conversion taken into account 
\citep[see details in][]{2009A&A...500..891S}
and  such thin atmospheres above condensed iron surface.  They also created a code
for modeling integral phase resolved spectra and light curves of rotating
 neutron stars. This code takes into account general
 relativistic effects and allows one to consider various temperature and
 magnetic field distributions. Analytical approximations of the
 considered three types of local spectra were used for such
 modeling. It was qualitatively shown that only a thin model
 atmosphere above a condensed iron surface can explain the observed
 equivalent width of the absorption feature and the pulsed
 fraction.

\begin{figure}[t] 
\resizebox{\hsize}{!}{ 
\includegraphics[angle=90,clip=]{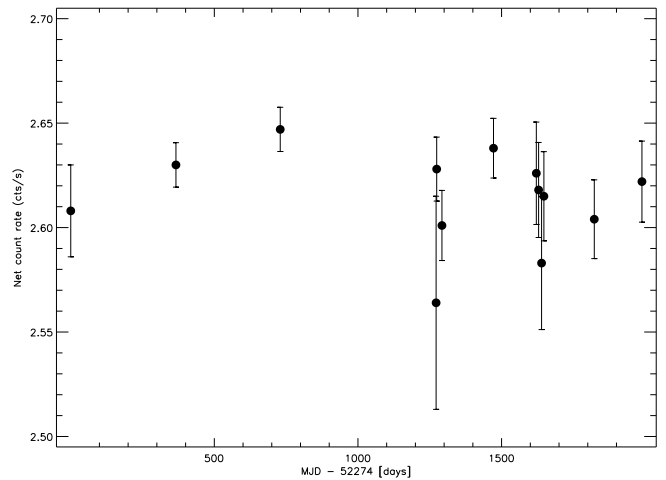}} 
\caption{Observed count rates of RBS1223 in the
energy band 0.16 -- 2.0\,keV in different \emph{XMM-Newton} EPIC pn observations. 
} 
\label{obs_ctr} 
\end{figure}

In this paper we use the code developed in Paper I to perform a comprehensive
study of co-added high-S/N and phase-resolved
{\em XMM-Newton} EPIC pn spectrum of RBS 1223, despite a possible small variations 
of brightness shown by this INS (see Fig.~\ref{obs_ctr}). 
 
\section{Observations and data reduction} 
 
RBS 1223 has been observed many times by \emph{XMM-Newton} (Table~\ref{obslog}). Here 
we focus on the data collected with EPIC pn  
\citep{2001A&A...365L...7D} from the 12 publicly available  
observations, with same instrumental setup (Full Frame mode, Thin1 filter, positioned on axis),  
in total presenting about 175~ks of effective exposure time. 
 
\begin{figure}[t] 
\resizebox{\hsize}{!}{ 
\includegraphics[clip=]{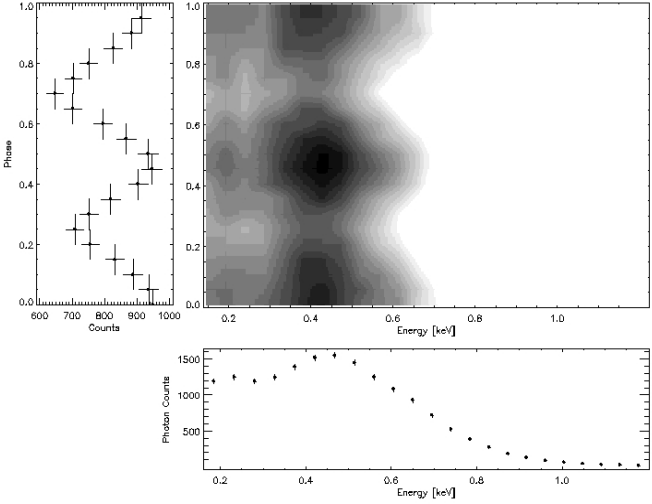}} 
\caption{Energy-Phase image of RBS1223 combined from different observations. Rotational phase-folded  
light curve in the broad energy, 0.2-2.0~keV, band (left panel) and the phase averaged spectrum (bottom panel) 
are shown. 
} 
\label{phaseener} 
\end{figure}

\begin{table} 
\caption[]{\emph{XMM-Newton} EPIC pn observations of  RBS1223} 
\label{obslog}      % is used to refer this table in the text 
\centering                          % used for centering table 
\begin{tabular}{cccc} 
\hline\noalign{\smallskip} 
Obs. ID & Obs. Date  & Exposure & Effective exposure \\ 
        &  MJD       &  ksec    &    ksec            \\ 
\hline 
0157360101 & 52640.4271023 & 28.910   &  25.966 \\ 
0163560101 & 53003.4674787 & 32.114   &  26.436 \\ 
0305900201 & 53546.0889925 & 16.806   &  11.443 \\ 
0305900301 & 53548.0919318 & 14.812   &  12.682 \\ 
0305900401 & 53566.4161765 & 14.814   &  10.574 \\ 
0305900601 & 53745.8803029 & 16.845   &  14.702 \\ 
0402850301 & 53894.9783016 &  7.419   &   4.761 \\ 
0402850401 & 53902.9489095 &  8.421   &   5.558 \\ 
0402850501 & 53913.2292603 & 12.517   &   2.801 \\ 
0402850701 & 53921.2857528 & 10.411   &   6.366 \\ 
0402850901 & 54096.6780925 &  9.315   &   8.011 \\ 
0402851001 & 54262.6501350 & 10.920   &   8.462 \\ 
\hline                                   %inserts single line 
\end{tabular} 
\end{table} 
 
The data were reduced using  standard threads from the {\em XMM-Newton}~data analysis  
package SAS version 10.0.0. We reprocessed all publicly available data  
(see Table~\ref{obslog}) with the standard metatask \emph{epchain}.  
To determine good time intervals free of background flares, we applied filtering  
expression on the background light curves, performing visual inspection. 
This reduced the total exposure time by $\sim$\,30\%.  
Solar barycenter corrected source and background photon events files  
and spectra were produced from the cleaned \emph{SINGLE}\footnote{See \emph{XMM-Newton} Users Handbook.} 
events, using an extraction radius of 30$\arcsec$  in all pointed observations. 
We extracted also light curves of RBS 1223 and corresponding backgrounds  
from nearby, source free regions. We then used the SAS task \emph{epiclccorr}  
to correct observed count rates for various sorts of detector inefficiencies  
(vignetting, bad pixels, dead time, effective areas, etc.) in different energy bands  
for each pointed observation (see Fig.~\ref{obs_ctr}). 
 
For each registered photon the corresponding rotational phase  
was computed according to the phase coherent timing solution  
provided by \cite{2005ApJ...635L..65K}. We produced spin phase-folded light curves  
in different energy bands and spectra corresponding to different phase intervals  
for each pointed observation. Finally, the latter ones were  
co-added for the same intervals of phases to create the combined phase-resolved spectra  
of RBS 1223 (Fig.~\ref{phaseener}). 
 
It should be noted that spectral responses and effective areas for those 12  
different observations were almost undistinguishable.     
 
\section{Data analysis}\label{da}
 
\begin{figure*} 
\includegraphics[width=16.5cm,clip=]{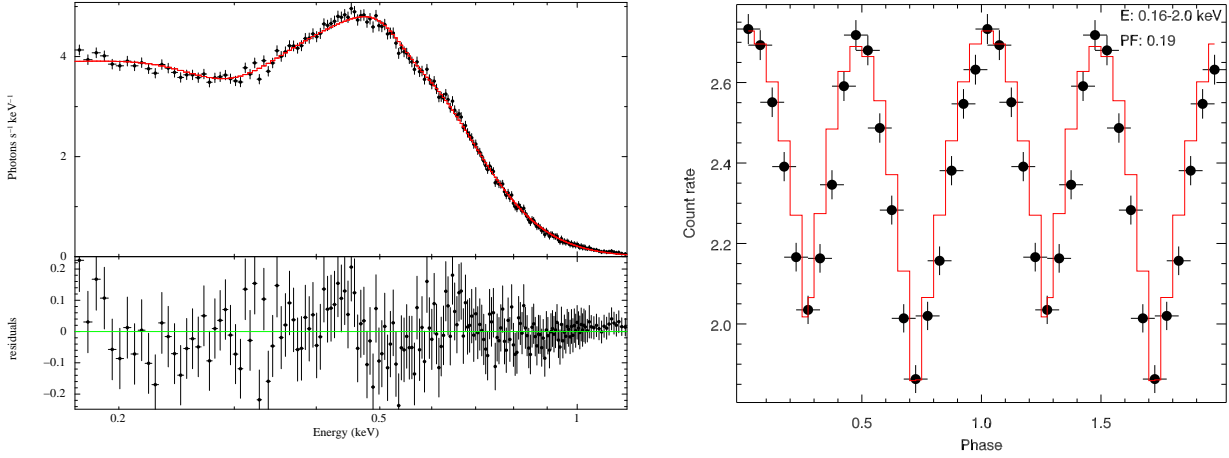} 
\caption{Phase-averaged X-ray spectrum (left panel) and phase-folded  
light curve (right panel) in the broad energy band of 0.16-2.0 keV of  
RBS 1223 combined from 12 pointed \emph{XMM-Newton} EPIC pn observations.  
} 
\label{gts1} 
\end{figure*}

First, we have fitted the   
phase-averaged,  high signal-to-noise ratio spectral data,
collected from different observations (Fig.~\ref{gts1} and Table~\ref{tgabs}),  
using a combination of an absorbed blackbody and a Gaussian absorption line  multiplicative component  
model (in {\it XSPEC phabs*bbodyrad*gabs}).  
 
\begin{table*} 
\centering                          % used for centering table 
\caption{Fitting results by model phabs*gabs*bbodyrad ($\tau=0.612\pm 0.08, \sigma=0.155\pm 0.008, N_{H}=(1.8\pm 0.0035)\times10^{18} cm^{-2}$) .  
\label{tgabs}}  
\begin{tabular}{clll} 
 Phase interval & kT     & Line center & Remarks \\ 
                & (keV)  & (keV)       &  \\ 
\hline 
0.20$-$0.35           &    0.084$\pm$0.001 &  0.25$\pm$0.01    &  First minimum  \\ 
0.35$-$0.65           &    0.088$\pm$0.001 &  0.26$\pm$0.01    &  Secondary peak \\ 
0.65$-$0.80           &    0.083$\pm$0.001 &  0.24$\pm$0.01    &  Second minimum \\ 
0.0$-$0.2,0.8$-$1.0   &    0.088$\pm$0.001 &  0.29$\pm$0.01    &  Primary peak  \\ 
0.00$-$1.0            &    0.086$\pm$0.001 &  0.27$\pm$0.01    &  Phase averaged \\ 
\hline 
\end{tabular} 
\end{table*} 
 
In the spectral energy range of 0.16$-$2.0~keV we obtained a statistically  
acceptable fit with the reduced $\chi^{2}=1.1$.  
 
\begin{figure*} 
\includegraphics[width=16.5cm,clip=]{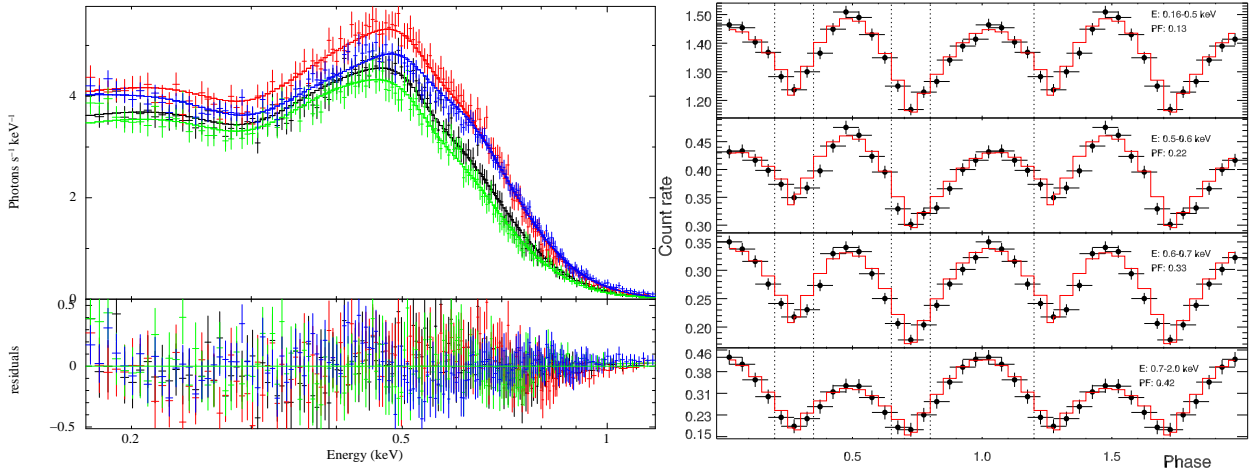} 
\caption{X-ray spectra including primary and secondary peaks, first and second  
minima, and phase-folded light curves in different energy bands of RBS 1223 combined from  
12 pointed \emph{XMM-Newton} EPIC pn observations. Fitted absorbed blackbody  
with gaussian absorption line models ({\it phabs*bbodyrad*gabs XSPEC}) to the spectra and  
two sinusoids to the phase-folded light curves are also shown (for details, see text and Table~\ref{tgabs}).} 
Dotted vertical lines are indicating phase intervals used for extraction and fitting of spectra shown in the left panel.
\label{gts2} 
\end{figure*}

We performed also a fit with the abovementioned spectral model 
for four phase intervals, including maxima and minima of the double-humped  
phase folded light curve, simultaneously. The fitted parameters are presented 
in Table~\ref{tgabs}. It shows a dependence of the blackbody apparent temperature 
and the Gaussian
absorption feature center upon rotational phase.
Next, we divided the broad energy band into four energy  
regions (0.16--0.5, 0.5--0.6, 0.6--0.7, and 0.7--2.0 keV) 
and constructed rotational phase folded light curves (Fig.~\ref{gts2}). 
It is also noteworthy that the higher the considered spectral energy range,  
the larger the pulsed fraction. These spin phase-folded light curves may serve 
for rough estimates of the viewing geometry and physical characteristics  
of emitting areas of RBS 1223 (see below). In particular, the obtained
  parameters of the fits at different phases (Table~\ref{tgabs}) support a
  model with hot areas around the magnetic poles, because the energies of line
  centers and the blackbody temperatures are larger at the peaks.
 
In view of this result, we implemented into the X-ray spectral  
fitting package {\it XSPEC} a number of 
new highly magnetized INS surface/atmosphere models developed in Paper~I.
They are based on various local models and compute rotational phase dependent integral emergent spectra of INS, 
using analytical approximations. The basic model includes 
temperature/magnetic field distributions over INS surface,  
viewing geometry and gravitational redshift. Three local
  radiating surface models are also considered, namely, 
a naked condensed iron surface and partially ionized hydrogen model atmospheres,  
semi-infinite or finite on top of the condensed surface. Here we have reproduced 
the essential part of the basic model (for details and further references, see Paper I). 
 
To compute an integral spectrum, the model uses an analytical expression for the
  local spectra: a diluted blackbody spectrum 
for both semi-infinite and thin models of a magnetized atmosphere with one absorption feature:  
\beq
\label{h1} 
     I_\mathrm{ E}(\alpha) = D~ B_\mathrm{ E}(T)~\phi(\alpha)\,\exp(-\tau), 
\eeq
where  
$\alpha$ is the angle between radiation propagation direction and the 
surface normal, and $\phi(\alpha)$ represents the considered angular 
distributions of the specific intensities for different 
cases of atmosphere models. We have represented $\phi(\alpha)$ by the following three models:
\be
\phi(\alpha)= 
\left\{ 
\begin{array}{ll} 
\label{h2} 
    1,  & \mathrm{\,a}, \\ 
    0.4215+0.86775 \cos\alpha, & \mathrm{\,b}, \\ 
   \left\{ 
     \begin{array}{l} \displaystyle
        \frac{1-0.2\cos^2\alpha_\mathrm{ c}}{1-\cos^2\alpha_\mathrm{ c}},
          \quad \textrm{if~~}
               \left\{
             \begin{array}{l}
                \cos\alpha \ge \cos\alpha_\mathrm{ c}
              \\ \textrm{and~~} E_\mathrm{c,i} < E < 4E_\mathrm{C},
             \end{array}
              \right.
         \\ 
        0.2, 
         \quad \textrm{if}~~~ \cos\alpha < \cos\alpha_\mathrm{ c}
          \textrm{~~and~~} E_\mathrm{c,i} < E < 4E_\mathrm{C},
         \\ 
        1 \quad \textrm{for other\, energies}.
    \end{array} 
   \right\}  
     & \mathrm{\,c}. \\ 
\end{array} 
\right.  
\ee 
Here, the cases a, b, and c correspond to the models of isotropic
emission, electron scattering directivity pattern,
 and a finite hydrogen atmosphere layer 
above a condensed iron surface, respectively;
$\cos\alpha_\mathrm{ c}= \left[\frac{1}{3}\left(E_\mathrm{ C_\mathrm{ 1}}/E_\mathrm{ C}-1\right)\right]^{2/3}$
is a parameter of angular distribution for the case of a thin atmosphere
in the energy range $E_\mathrm{ci}<E<4E_\mathrm{C}$;
$E_\mathrm{ c,i}=\hbar~ZeB/m_\mathrm{ i}c$ and $E_\mathrm{ c,e}=\hbar eB/m_\mathrm{ e}c$ 
are the ion and electron cyclotron energies, 
$E_\mathrm{ p,e}=\hbar \sqrt{4\pi e^2 n_\mathrm{ e}/m_\mathrm{ e}}$ is the electron plasma energy,
$Z$ and $m_\mathrm{ i}$ are the ion charge number and ion mass, 
$n_\mathrm{ e}$ is the electron number density, $B$ the magnetic field strength, 
\beq
 \label{h3} 
E_\mathrm{ C} = E_\mathrm{ c,i} + E_\mathrm{ p,e}^2/E_\mathrm{ c,e}, 
\eeq
and 
\beq
\label{h4}
   E_\mathrm{ C_1} = E_\mathrm{ C}~(1+3(1-\cos\alpha)^{3/2}).
\eeq
 
The optical depths $\tau = \tau^{0} \exp\left(-\frac{(E-E_\mathrm{
    line})^2}{2\sigma_\mathrm{ line}^2}\right)$  
and the widths $\sigma_\mathrm{ line}$ of the absorption features 
 are considered 
identical for all local spectra, but the center of the line depends on the local
  magnetic field strength.  In the case of a semi-infinite atmosphere this absorption
  line represents the blend of the proton cyclotron line and nearby b-b
  atomic transitions in neutral hydrogen. The line center corresponds to the
  proton cyclotron energy
  $E_\mathrm{ c,H}$. In the case a thin atmosphere the
  absorption feature was represented by a half of a Gaussian line. It means that
  $\tau^0 = 0$ at $E < E_\mathrm{ line}$. This half of the Gaussian line represents a
the  complex absorption feature, which includes a broad absorption feature from
  the emitting condensed iron surface transmitted through a thin hydrogen
  atmosphere and  the proton cyclotron line plus  b-b
  atomic hydrogen transitions (see details in Paper I). The center of this
  line corresponds to the ion cyclotron $E_\mathrm{ c,i}$ for completely
  ionized iron. In both cases, the
  absorption details are better represented by sums of two or even three Gaussian
  lines. But we have chosen to use only one (or half) Gaussian line to reduce the
number of fitting parameters.
The local spectra of the condensed iron surface is
approximated by simple step functions (see Paper I). 
The dilution factor $D$ cannot be found from the fitting independently, and we
take $D=1$ for all local spectra.  

The blackbody temperature and the magnetic field  distributions in the two
  emitting areas around the magnetic poles were presented analytically by \cite{2006A&A...451.1009P}:  
\beq
\label{h5}
T^4 = T_\mathrm{ p1,2}^4 \frac{\cos^2\theta}{\cos^2\theta + a_\mathrm{ 1,2}\sin^2\theta } + T_\mathrm{ min}^4 
\eeq
and  
\beq
\label{h6}
B = B_\mathrm{ p1,2} \sqrt{\cos^2\theta + a_{1,2}\sin^2\theta},  
\eeq
where  $T_\mathrm{ p1,2}$ and $B_\mathrm{ p1,2}$ are the temperatures and magnetic field strengths  
at the poles and $a_\mathrm{ 1,2}$ their  distribution 
parameters and $T_\mathrm{ min}$ is the minimum temperature reached on the surface of the star  
\citep[we chose $\approx 0.3T_\mathrm{ p}$, like in][]{2006A&A...451.1009P}. Here
$\theta$ is the magnetic colatitude. 
 
The parameters $a_\mathrm{ 1,2}$ are approximately equal to the squared ratio of the magnetic
field strength at the equator to the field strength at the pole,
$a_\mathrm{ 1,2} \approx (B_\mathrm{ eq}/B_\mathrm{ p1,2})^2 $.
Using these parameters we can describe various temperature distributions, from strongly 
peaked ($a \gg 1$) to the classical dipolar ($a = 1/4$)  
and homogeneous ($a = 0$) ones.
 
Finally, the total observed flux at distance $d$ from the INS is summed
  over the visible surface. 
A local spectrum at photon energy $E$ from the
  unit surface is computed taking into account viewing geometry,
  gravitational redshift, and light  
bending effects \citep[e.g.,][]{2003MNRAS.343.1301P}:
 \beq
 \label{h7} 
 f_{\mathrm{E}}
\,d\,\varphi\,d\gamma = C_\mathrm{norm}\,
(1+z)^{-3}\, 
I_{\mathrm{E'}}(\alpha)\,\cos\alpha\, \sin\gamma \,d\,\varphi\,d\gamma, 
\eeq
where $C_\mathrm{norm}=R^2/d^2$ is a normalization constant and
$\gamma$ is the angle between the radius vector at a 
given point and the rotation axis. The observed and emitted photon energies are
related as $E=~E'(1+z)$, where $z$ is gravitational redshift.
The angle $\alpha$ between the emitted photon and normal to the surface 
in the local reference frame depends
on $z$, the local surface position (on azimuthal and colatitude angles
$\varphi$ and $\gamma$),
 and the angle between rotation axis and line of sight $i$, and is computed
 using the approximation of \citet{2002ApJ...566L..85B}.
The summation is performed separately for both regions around magnetic
poles, because we allow the poles to be not precisely antipodal (one of
them can be shifted by angle $\kappa$ relative to the symmetric antipodal
position). Therefore the poles may
have different angles between rotation and magnetic axes
$\theta_\mathrm{ B1,2}$.
 
As shown in Paper I, a model with a thin hydrogen 
atmosphere above the condensed iron surface with a smooth 
temperature distribution over the neutron star surface 
may describe very well the observed physical properties of RBS 1223. 
 
This basic model has a number of input parameters depending on  
the inner atmosphere boundary condition of an INS (condensed iron surface or blackbody, 
temperature and magnetic field distributions over surface, viewing geometry and gravitational redshift) 
and angular distribution of the emergent radiation (isotropic or peaked by
a thin atmosphere above a condensed surface). 

Combined spectra of RBS 1223 in 20 phase bins (Fig.~\ref{phaseener}) were considered as  
different data sets during simultaneous fitting with the absorbed abovementioned model.  
All input parameters were free and linked between those data groups. 
 
Given the large number of the free parameters in the model we performed a preliminary analysis 
in order to get rough estimates of (or constraints on) some of them. From the observed  
double-humped light curve shape in different energy bands (see Fig.~\ref{gts2}),  
it has been already clear that two emitting areas have different spectral and geometrical  
characteristics (e.g., the relatively cooler one has a larger size). 
Moreover, from the peak-to-peak separation in this double-humped light  
curve we locate the cooler one at an offset angle of $\kappa$ with respect to the  
magnetic axis and azimuth \citep{2005A&A...441..597S}. 
 
First we performed formal spectral fitting  
with the simplest model, i.e.\ absorbed blackbody with multiplicative gaussian absorption line  
including phases of maxima (see Table~\ref{tgabs}) in the  light curve (Figs.~\ref{gts1} and \ref{gts2}).  
It is clear, that at these phases X-ray emission is mostly dominated by emitting areas at  
the magnetic poles, where temperatures also have maxima  
(see formula of temperature and magnetic field dependence upon polar angle,
 \citealp{2006A&A...451.1009P} and Paper I). 
Secondly, some constraints on magnetic field strengths at the poles may be used on the base of  
period and its derivative values assuming magnetic dipole breaking as a  
main mechanism of the spin down of RBS 1223. 
 
Having these two general constraints on temperatures and  
magnetic field strengths at the poles we simulated a large number 
of photon spectra (absorbed blackbody with gaussian absorption line and 
different models considered in the Paper I) folded with the response of 
{\em XMM-Newton} EPIC pn camera, taking into account also the 
interstellar absorption (for parameters see Table~\ref{tgabs}, 
and using a characteristic magnetic field strength of B$\sim$~3.4$\times$10$^{13}$ Gauss).

The predicted phase-folded light curves in four spectral 
ranges: 0.16-0.5~keV, 0.5-0.6~keV, 0.6-0.7~keV, 0.7-2.0~keV
and with free parameters of viewing geometry and 
gravitational redshift, normalized to the maximum of the brightness, 
we cross-correlated with observed ones (Fig.~\ref{gts2}).
We infer some constraints on the parameters from the unimodal  
distributions of them when cross-correlation coefficients were exceeding 0.9 in  
the mentioned four energy bands simultaneously and used them as  
initial input value and as lower and upper bounds for fitting purposes.  
For example, gravitational redshift cannot exceed 0.3 (it is not possible
to obtain the observed PF at larger $z$ due to strong light bending) or the antipodal shift angle  
must be less than $25^\circ$ (due to the observed phase separation between the two peaks
  in the light curve).  
It is also evident that the sum of the inclination angle 
of the line of sight and magnetic poles relative to the rotational axis are already  
constrained by the light curve class \citep[see,][class III]{2006MNRAS.373..836P} 
 and have the maximum effect (e.g., provide the maximum PF) when both are equal to $90^\circ$. 
It should be noted that these angles do not have a large influence on the fitting,
and the only important issue is a range of values when two maxima are observed. 
 
Having abovementioned crude constraints and input values of free parameters we performed fitting  
with the models implemented in {\it XSPEC} package of combined spectra of RBS 1223 in 20   
phase bins (Fig.~\ref{phaseener}) simultaneously, i.e. each of those phase resolved   
spectrum considered as different data sets with the linked parameters to the others,  
and the only differences were phase ranges, which were fixed for an individual phase resolved spectrum.  
 
The fitting was successful, with C-statistic value 2937 with 2159  
degrees of freedom. These parameters are presented in Table~\ref{fit1}. 
 
\begin{table*} 
%\centering                          % used for centering table 
\caption[]{Simultaneous fitting results of combined, phase resolved X-ray spectra of RBS1223 with different spectral models} 
\label{fit1} 
\centering                          % used for centering table 
\begin{tabular}{lcc} 
\hline\noalign{\smallskip} 
Fitted & \multicolumn{2}{c}{Spectral Model} \\ 
\cline{2-3} 
Parameter\tablefootmark{*} & Iron condensed surface & Blackbody \\ 
\cline{2-3} 
          & partially ionized H atmosphere & electron scattering \\  
\hline\hline 
 $T_\mathrm{ p1}~\mathrm{ [eV]}$          & $\mathrm{ 105.0_{-4.0}^{+2.0}}$      & $109 \pm 4.0$    \\    
 $T_\mathrm{ p2}~\mathrm{ [eV]} $         &  $99.0 \pm 3.0$                & $106 \pm 3.0$    \\  
 $B_\mathrm{ p1}\times 10^{14}~\mathrm{ [G]}$  & $0.86 \pm 0.02$                & $0.65 \pm 0.03$   \\ 
 $B_\mathrm{ p2}\times 10^{14}~\mathrm{ [G]}$  & $0.86 \pm 0.02$                & $0.58 \pm 0.02$  \\ 
 $a_\mathrm{ p1}$                   & $0.61 \pm 0.11$                & 0.25         \\ 
 $a_\mathrm{ p2}$                   & $0.29 \pm 0.05$                & 0.25        \\  
 $\tau^0$                        & $\mathrm{ 2.76_{-0.02}^{+0.10}}$      & $1.90 \pm 0.06$  \\ 
 $\sigma~\mathrm{ [eV]}$               & $\mathrm{ 225.8_{-1.8}^{+5.6}}$      & $168.0 \pm 6.0$  \\ 
 $z     $                        & $0.15 \pm 0.02$                & $0.17 \pm 0.03$  \\ 
 $\kappa[{\degr}]$                   & $4.2 \pm 0.6$                 & $4.0 \pm 0.4$    \\ 
 ${\it i[{\degr}]}$                  & $48.9 \pm 0.5$                 & $45.6 \pm 0.5$   \\ 
 $\theta[{\degr}]$                   & $90.0 \pm 0.5$                 & $90.0 \pm 0.5$   \\ 
\hline 
\end{tabular} 
\tablefoot{\tablefoottext{*}{See section~\ref{da} for definition of the model parameters}}
\end{table*} 
 
\begin{figure}[t] 
\resizebox{\hsize}{!}{ 
\includegraphics[clip=]{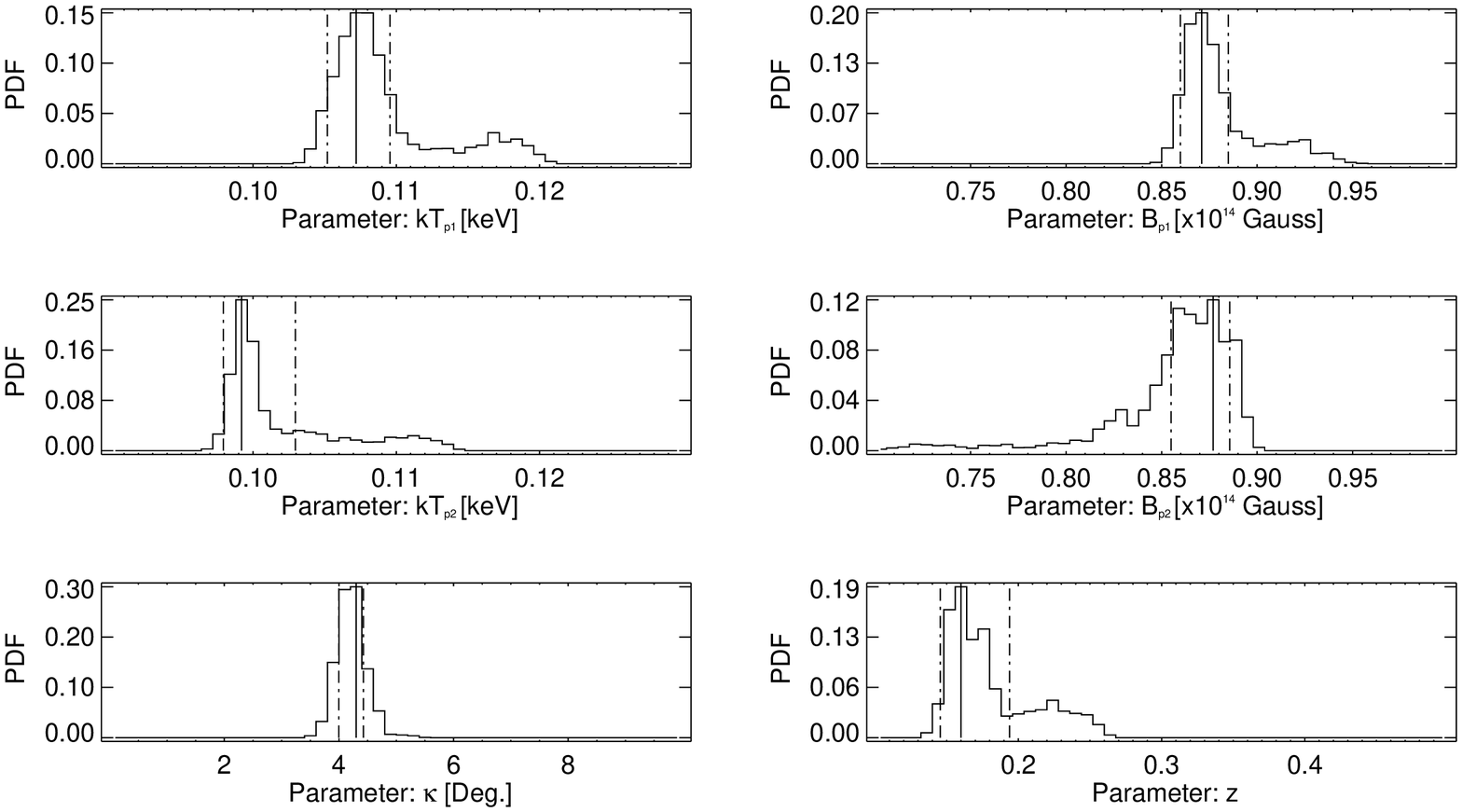}} 
\caption{Probability density distributions of parameters (temperatures, magnetic field strengths at the poles,  
antipodal shift angle and gravitational redshift) by MCMC fitting with the model of a neutron star with  
condensed surface and partially ionized hydrogen layer above it. The most
probable parameter value is indicated by the solid vertical line.  
Dashed vertical lines indicate the highest probability interval ($68\%$, for details see text). 
} 
\label{mcmc_res} 
%} 
\end{figure}

In order to assess a degree of uniqueness and to estimate confidence intervals  
of the determined parameters, we have additionally performed Markov Chain Monte Carlo (MCMC) fitting 
as implemented in {\it XSPEC}.  
In Fig.~\ref{mcmc_res} we present probability density distributions of some of them.  
Note, independent initial input values of parameters of  
MCMC approach converged to the same values, in 6 different chains.

\section{Discussion} 
 
The combined phase-resolved spectra of RBS 1223 can be simultaneously fitted by emergent radiation 
of a spectral model of an iron condensed surface with a partially ionized hydrogen atmosphere above it. 
Formally they can be fitted also by a blackbody spectrum with proton-cyclotron absorption  
gaussian line and a peaked (typical for an electron scattering atmosphere) angular distribution of the emergent
  radiation. In both cases two emitting areas with slightly different  
characteristics are required (see Table~\ref{fit1}). 
 
It is worth to note, that the resulting fit parameters are very similar for different
spectral models (see Table~\ref{fit1}), which is also confirmed by an MCMC approach with different input parameters. 
 
However, we believe that the emission properties due  the condensed surface
model with a partially ionized,  
optically thin hydrogen layer above it, including vacuum polarization effects, 
is more physically motivated.
Moreover, semi-infinite atmospheres have rather
fan-beamed emergent radiation (see Paper I and references within) and it seems 
impossible to combine a proton cyclotron line with a pencil-beamed emergent radiation.
 
\begin{figure}[t] 
\resizebox{\hsize}{!}{ 
\includegraphics[angle=90,clip=]{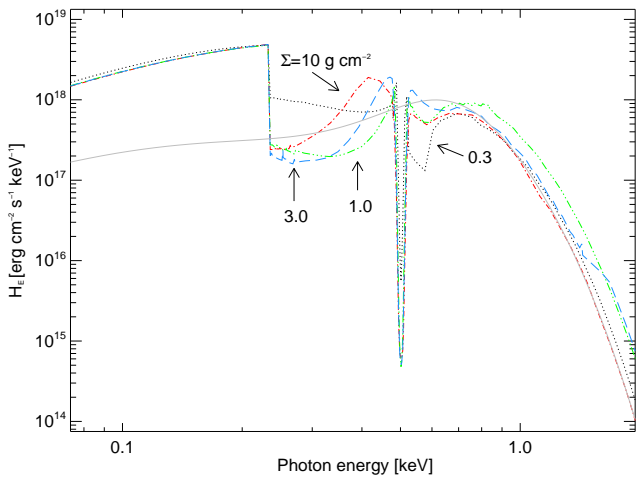}} 
\caption{Emergent spectra of the fitted model of a magnetized atmosphere with a
  condensed iron surface and a partially ionized 
hydrogen layer above it (see Table~\ref{fit1}). Different lines are
corresponding different atmospheric surface densities. 
For comparison purposes a blackbody spectrum with gaussian absorption line is
also shown in gray (see text). 
} 
\label{vs_model} 
\end{figure}

We calculated a set of thin highly magnetized partially ionized hydrogen
atmospheres above a condensed iron surface with magnetic field strength
$B=8\times 10^{13}$ G, which is close to the value
estimated from observations. The observed blackbody
temperature of the spectra is reproduced at effective
temperatures $T_\mathrm{ eff} \approx 7 \times 10^5$~K. Examples of the computed
emergent spectra with parameters $B=8\times 10^{13}$~G, $T_\mathrm{ eff} = 7 \times
10^5$~K and various atmosphere column densities $\Sigma$ are shown in
Fig.~\ref{vs_model}. The emergent diluted blackbody spectrum
with $T=0.1$~keV and the absorption Gaussian line parameters, presented in
Table~\ref{fit1} ($\tau^0=2.8, \sigma_\mathrm{ line}=0.226$~keV, $E_\mathrm{ line} = 0.24$~keV)
and $D=0.34$~ are also shown. Unfortunately, from these models we cannot
evaluate the actual atmosphere thickness (the surface density), but we can
obtain the dilution factor $D$, which is important to correct the distance estimation.

Emission spectra based on realistic temperature and magnetic field  
distributions with strongly magnetized hydrogen atmospheres (or other light elements)  
are formally still an alternative\footnote{Our attempt to fit the combined,  
phase-averaged spectrum of RBS 1223 by  partially ionized, strongly  
magnetized hydrogen or mid-Z element plasma model  
\citep[{\it XSPEC nsmax},][]{2007MNRAS.377..905M,2008ApJS..178..102H},  
as well two spots or purely condensed iron surface models, failed. 
Noteworthy, an acceptable fit is obtained by {\it nsmax} model with  
additional, multiplicative gaussian absorption line component (model {\it
  gabs}).},  but  it is unphysical because of the absorption line added by hand.
  
A purely proton-cyclotron absorption line scenario  
can be excluded owing to the equivalent width of the observed absorption spectral feature  
%(e.g., Doppler smearing and magnetic broadening) 
in the X-ray spectrum of RBS 1223. Magnetized semi-infinite atmospheres
  predict too low an equivalent width of the proton cyclotron line in comparison
  with the observed one.
 
\begin{figure}[t] 
\resizebox{\hsize}{!}{ 
\includegraphics[clip=]{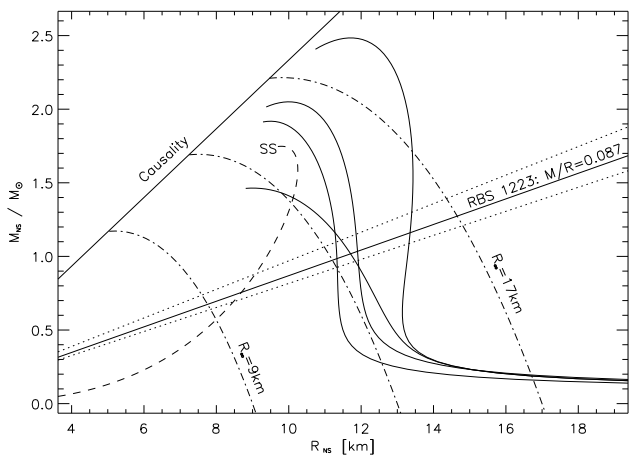}} 
\caption{Mass-radius relations for several equations of state \citep[thin solid 
curves,][]{2007ASSL..326.....H}, 
and a strange star (dashed).  Thin dash-doted: curves of 
constant apparent radius $R_{\infty}=R/\sqrt{1-2GM/Rc^2}=$~9,13 and 17 km  
\citep{2005esns.conf..117T}. } 
\label{figmr} 
\end{figure} 
 
This result of the fitting (with the condensed surface model with partially ionized,  
optically thin hydrogen atmosphere above it, including vacuum polarization effects)  
suggests a true radius of RBS 1223 of $16 \pm 1\mathrm{km}$ for a standard
neutron star of 1.4 solar mass,  
considerably larger than the canonical radius of 10 km; 
it is only marginally compatible with the range from $\approx
10$ km to $\approx 14$ km, allowed by modern theoretical equations
of state of superdense matter 
(\citealp{2007ASSL..326.....H}; \citealp[][and references therein]{2010PhRvL.105p1102H}),
and indicates a very stiff equation of state of RBS 1223 (Fig.~\ref{figmr};
\citealp[for similar results, see also][]{2007MNRAS.375..821H,2006ApJ...644.1090H,2006MNRAS.369.2036S,2010arXiv1004.4871S}).

With these estimates of the radius and normalization constant of the fit ($R/d
\approx 0.0247$~km~pc$^{-1}$), we obtained for the assessment of the distance
of RBS 1223 $\sqrt{D}\, 650_{-50}^{+25} \textrm{~pc} = \sqrt{D/0.34}\,
380_{-30}^{+15}$~pc \citep[see also][]{2005A&A...441..597S}.
 
\section{Conclusions} 
 
The observed phase resolved spectra of the INS RBS 1223   
are satisfactorily fitted with two slightly different physical and geometrical  
characteristics of emitting areas, by 
the model of a condensed iron surface, with partially ionized,  
optically thin hydrogen atmosphere above it, including vacuum polarization effects,  
as orthogonal rotator. The fit also suggests the absence of a strong toroidal  
magnetic field component. Moreover, the determined mass-radius ratio
($(M/M_{\sun})/(R/\mathrm{km})=0.087 \pm 0.004 $) 
suggests a very stiff equation of state of RBS 1223. 
 
These results on RBS 1223 are promising, since we could find good simultaneous  
fits to the rotational phase-resolved spectra of RBS 1223 
using analytic approximations for the abovementioned model implementation. 
 
More work for detailed spectral model computation will be certainly worth to do in the near future 
and application to the phase-resolved spectra of other INSs. In particular, 
for RX J1856.5$-$3754 and RX J0720.4$-$3125, including high resolution spectra observed by
\emph{XMM-Newton}
and \emph{Chandra}
with possibly other absorption features \citep{2009A&A...497L...9H,2010A&A...518A..24P}. 
 
\begin{acknowledgements} 
      VH and VS acknowledge support by the German 
      \emph{Deut\-sche For\-schungs\-ge\-mein\-schaft (DFG)\/} through project 
      C7 of SFB/TR~7 ``Gravitationswellenastronomie''. A.Y.P. acknowledges  
     partial support from the RFBR (Grant 11-02-00253-a) 
     and the Russian Leading Scientific Schools program (Grant NSh-3769.2010.2). 
 
\end{acknowledgements} 
 
\bibliographystyle{aa} % style aa.bst 
\bibliography{hambaryanetal} % your references Yourfile.bib 

\begin{thebibliography}{29}
\expandafter\ifx\csname natexlab\endcsname\relax\def\natexlab#1{#1}\fi

\bibitem[{{Beloborodov}(2002)}]{2002ApJ...566L..85B}
{Beloborodov}, A.~M. 2002, \apjl, 566, L85

\bibitem[{{den Herder} {et~al.}(2001){den Herder}, {Brinkman}, {Kahn},
  {Branduardi-Raymont}, {Thomsen}, {Aarts}, {Audard}, {Bixler}, {den Boggende},
  {Cottam}, {Decker}, {Dubbeldam}, {Erd}, {Goulooze}, {G{\"u}del}, {Guttridge},
  {Hailey}, {Janabi}, {Kaastra}, {de Korte}, {van Leeuwen}, {Mauche},
  {McCalden}, {Mewe}, {Naber}, {Paerels}, {Peterson}, {Rasmussen}, {Rees},
  {Sakelliou}, {Sako}, {Spodek}, {Stern}, {Tamura}, {Tandy}, {de Vries},
  {Welch}, \& {Zehnder}}]{2001A&A...365L...7D}
{den Herder}, J.~W., {Brinkman}, A.~C., {Kahn}, S.~M., {et~al.} 2001, \aap,
  365, L7

\bibitem[{{Haberl}(2007)}]{2007Ap&SS.308..181H}
{Haberl}, F. 2007, \apss, 308, 181

\bibitem[{{Haensel} {et~al.}(2007){Haensel}, {Potekhin}, \&
  {Yakovlev}}]{2007ASSL..326.....H}
{Haensel}, P., {Potekhin}, A.~Y., \& {Yakovlev}, D.~G. 2007, Astrophysics and
  Space Science Library, Vol. 326, {Neutron Stars 1: Equation of State and
  Structure} (Springer, New York)

\bibitem[{{Hambaryan} {et~al.}(2009){Hambaryan}, {Neuh{\"a}user}, {Haberl},
  {Hohle}, \& {Schwope}}]{2009A&A...497L...9H}
{Hambaryan}, V., {Neuh{\"a}user}, R., {Haberl}, F., {Hohle}, M.~M., \&
  {Schwope}, A.~D. 2009, \aap, 497, L9

\bibitem[{{Hebeler} {et~al.}(2010){Hebeler}, {Lattimer}, {Pethick}, \&
  {Schwenk}}]{2010PhRvL.105p1102H}
{Hebeler}, K., {Lattimer}, J.~M., {Pethick}, C.~J., \& {Schwenk}, A. 2010,
  Physical Review Letters, 105, 161102

\bibitem[{{Heinke} {et~al.}(2006){Heinke}, {Rybicki}, {Narayan}, \&
  {Grindlay}}]{2006ApJ...644.1090H}
{Heinke}, C.~O., {Rybicki}, G.~B., {Narayan}, R., \& {Grindlay}, J.~E. 2006,
  \apj, 644, 1090

\bibitem[{{Ho} {et~al.}(2007){Ho}, {Kaplan}, {Chang}, {van Adelsberg}, \&
  {Potekhin}}]{2007MNRAS.375..821H}
{Ho}, W.~C.~G., {Kaplan}, D.~L., {Chang}, P., {van Adelsberg}, M., \&
  {Potekhin}, A.~Y. 2007, \mnras, 375, 821

\bibitem[{{Ho} {et~al.}(2008){Ho}, {Potekhin}, \&
  {Chabrier}}]{2008ApJS..178..102H}
{Ho}, W.~C.~G., {Potekhin}, A.~Y., \& {Chabrier}, G. 2008, \apjs, 178, 102

\bibitem[{{Ho} {et~al.}(2009){Ho}, {Potekhin}, {Chabrier}, \&
  {Mori}}]{2009AAS...21343609H}
{Ho}, W.~C.~G., {Potekhin}, A.~Y., {Chabrier}, G., \& {Mori}, K. 2009, in
  Bulletin of the American Astronomical Society, Vol.~41, Bulletin of the
  American Astronomical Society, 308

\bibitem[{{Kaplan} \& {van Kerkwijk}(2005)}]{2005ApJ...635L..65K}
{Kaplan}, D.~L. \& {van Kerkwijk}, M.~H. 2005, \apjl, 635, L65

\bibitem[{{Mereghetti}(2008)}]{2008A&ARv..15..225M}
{Mereghetti}, S. 2008, \aapr, 15, 225

\bibitem[{{Mori} \& {Ho}(2007)}]{2007MNRAS.377..905M}
{Mori}, K. \& {Ho}, W.~C.~G. 2007, \mnras, 377, 905

\bibitem[{{P{\'e}rez-Azor{\'{\i}}n}
  {et~al.}(2006{\natexlab{a}}){P{\'e}rez-Azor{\'{\i}}n}, {Miralles}, \&
  {Pons}}]{2006A&A...451.1009P}
{P{\'e}rez-Azor{\'{\i}}n}, J.~F., {Miralles}, J.~A., \& {Pons}, J.~A.
  2006{\natexlab{a}}, \aap, 451, 1009

\bibitem[{{P{\'e}rez-Azor{\'{\i}}n}
  {et~al.}(2006{\natexlab{b}}){P{\'e}rez-Azor{\'{\i}}n}, {Pons}, {Miralles}, \&
  {Miniutti}}]{2006A&A...459..175P}
{P{\'e}rez-Azor{\'{\i}}n}, J.~F., {Pons}, J.~A., {Miralles}, J.~A., \&
  {Miniutti}, G. 2006{\natexlab{b}}, \aap, 459, 175

\bibitem[{{Potekhin}(2010)}]{2010A&A...518A..24P}
{Potekhin}, A.~Y. 2010, \aap, 518, A24+

\bibitem[{{Poutanen} \& {Beloborodov}(2006)}]{2006MNRAS.373..836P}
{Poutanen}, J. \& {Beloborodov}, A.~M. 2006, \mnras, 373, 836

\bibitem[{{Poutanen} \& {Gierli{\'n}ski}(2003)}]{2003MNRAS.343.1301P}
{Poutanen}, J. \& {Gierli{\'n}ski}, M. 2003, \mnras, 343, 1301

\bibitem[{{Schwope} {et~al.}(2005){Schwope}, {Hambaryan}, {Haberl}, \&
  {Motch}}]{2005A&A...441..597S}
{Schwope}, A.~D., {Hambaryan}, V., {Haberl}, F., \& {Motch}, C. 2005, \aap,
  441, 597

\bibitem[{{Schwope} {et~al.}(2007){Schwope}, {Hambaryan}, {Haberl}, \&
  {Motch}}]{2007Ap&SS.308..619S}
{Schwope}, A.~D., {Hambaryan}, V., {Haberl}, F., \& {Motch}, C. 2007, \apss,
  308, 619

\bibitem[{{Schwope} {et~al.}(1999){Schwope}, {Hasinger}, {Schwarz}, {Haberl},
  \& {Schmidt}}]{1999A&A...341L..51S}
{Schwope}, A.~D., {Hasinger}, G., {Schwarz}, R., {Haberl}, F., \& {Schmidt}, M.
  1999, \aap, 341, L51

\bibitem[{{Suleimanov} {et~al.}(2010{\natexlab{a}}){Suleimanov}, {Hambaryan},
  {Potekhin}, {van Adelsberg}, {Neuh{\"a}user}, \&
  {Werner}}]{2010A&A...522A.111S}
{Suleimanov}, V., {Hambaryan}, V., {Potekhin}, A.~Y., {et~al.}
  2010{\natexlab{a}}, \aap, 522, A111+

\bibitem[{{Suleimanov} {et~al.}(2009){Suleimanov}, {Potekhin}, \&
  {Werner}}]{2009A&A...500..891S}
{Suleimanov}, V., {Potekhin}, A.~Y., \& {Werner}, K. 2009, \aap, 500, 891

\bibitem[{{Suleimanov} \& {Poutanen}(2006)}]{2006MNRAS.369.2036S}
{Suleimanov}, V. \& {Poutanen}, J. 2006, \mnras, 369, 2036

\bibitem[{{Suleimanov} {et~al.}(2010{\natexlab{b}}){Suleimanov}, {Poutanen},
  {Revnivtsev}, \& {Werner}}]{2010arXiv1004.4871S}
{Suleimanov}, V., {Poutanen}, J., {Revnivtsev}, M., \& {Werner}, K.
  2010{\natexlab{b}}, ArXiv e-prints

\bibitem[{{Tr{\"u}mper}(2005)}]{2005esns.conf..117T}
{Tr{\"u}mper}, J.~E. 2005, in NATO ASIB Proc. 210: The Electromagnetic Spectrum
  of Neutron Stars, ed. {A.~Baykal, S.~K.~Yerli, S.~C.~Inam, \& S.~Grebenev},
  117--132

\bibitem[{{Tr{\"u}mper} {et~al.}(2004){Tr{\"u}mper}, {Burwitz}, {Haberl}, \&
  {Zavlin}}]{2004NuPhS.132..560T}
{Tr{\"u}mper}, J.~E., {Burwitz}, V., {Haberl}, F., \& {Zavlin}, V.~E. 2004,
  Nuclear Physics B Proceedings Supplements, 132, 560

\bibitem[{{Turolla}(2009)}]{2009ASSL..357..141T}
{Turolla}, R. 2009, in Astrophysics and Space Science Library, Vol. 357,
  Neutron Stars and Pulsars, ed. {W.~Becker}, 141--164

\bibitem[{{van Adelsberg} \& {Lai}(2006)}]{2006MNRAS.373.1495V}
{van Adelsberg}, M. \& {Lai}, D. 2006, \mnras, 373, 1495

\end{thebibliography}
\end{document}